\documentclass{PoS}
\usepackage{graphicx}

\title{Direct determination of strange and light quark condensates from full lattice QCD}

\ShortTitle{Strange and light quark condensates from lattice QCD}

\author{\speaker{C. T. H. Davies}\\
          \\
        University of Glasgow\\
        E-mail: \email{Christine.Davies@glasgow.ac.uk}}

\author{C. McNeile\\
        Universit\"{a}t Wuppertal}
\author{A. Bazavov\\
        Brookhaven National Laboratory}
\author{R. J. Dowdall\\
        University of Glasgow}
\author{K. Hornbostel\\
        Southern Methodist University}
\author{G. P. Lepage\\
        Cornell University}
\author{H. Trottier\\
        Simon Fraser University}

\abstract{ We determine the strange and light quark condensates 
in full lattice QCD for the first time. This is done by direct 
calculation of the expectation value of the trace of the quark 
propagator followed by subtraction of the appropriate perturbative
contribution to convert to a value for the condensate in the $\overline{MS}$ 
scheme at 2 GeV. We use lattice QCD configurations including 
$u$, $d$, $s$ and $c$ quarks in the sea with $u/d$ quark masses 
going down to the physical value. We find the ratio of the 
strange to the light quark condensate to be 1.08(16). }

\FullConference{Xth Quark Confinement and the Hadron Spectrum,\\
		October 8-12, 2012\\
		TUM Campus Garching, Munich, Germany}

\begin{document}

\section{Introduction}
The condensation of quark-antiquark pairs in the vacuum 
signalling the breakdown of chiral symmetry is an important 
feature of low energy QCD. The value of the chiral condensate 
(the quark condensate for zero quark mass) is given by 
the Gell-Mann, Oakes, Renner relation~\cite{gmor}: 
\begin{equation}
\frac{f_{\pi}^2M_{\pi}^2}{4} = - \frac{m_u+m_d}{2}\frac{\langle 0 | \overline{u}u + \overline{d}d | 0 \rangle}{2}.
\label{eq:gmor} 
\end{equation}
A value for the chiral condensate can be extracted accurately 
but indirectly from chiral extrapolation of lattice QCD results 
for meson masses and decay constants. 
A harder question to answer is that of the value of the light and 
strange quark condensates at their physical non-zero quark masses. 
This is because the condensate must be carefully defined, taking 
into account the quark mass dependent mixing of $m\overline{\psi}\psi$ (not 
normal-ordered) with 
the unit operator. On the lattice, in a direct calculation of 
the condensate, such mixing gives rise to terms which diverge 
as $m/a^2$. We have calculated these terms, through $\mathcal{O}(\alpha_s)$  
in lattice QCD perturbation theory and used these to convert the 
lattice QCD results to a well-defined condensate in the $\overline{MS}$ 
scheme at a fixed scale (2 GeV). Since we have results at multiple 
values of the lattice spacing we can fit for remaining $m/a^2$ pieces 
at higher order in $\alpha_s$. We can also extrapolate our results 
to the continuum  limit at $a=0$. 
Since the conference we have finalised a paper on this work~\cite{cond} 
and refer the reader there for all details. 

\begin{figure}
\begin{center}
\includegraphics[width=0.4\hsize]{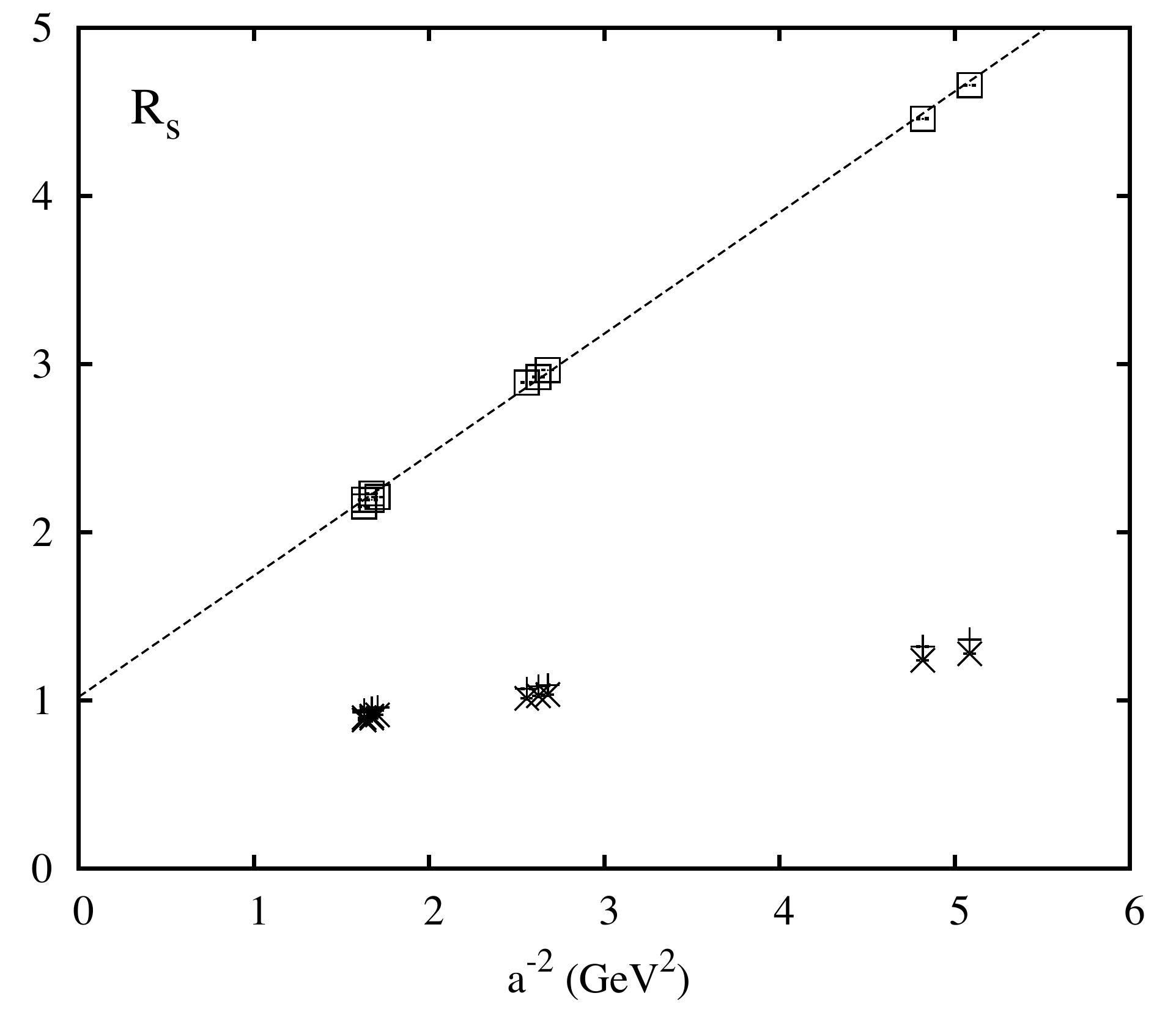}
\includegraphics[width=0.4\hsize]{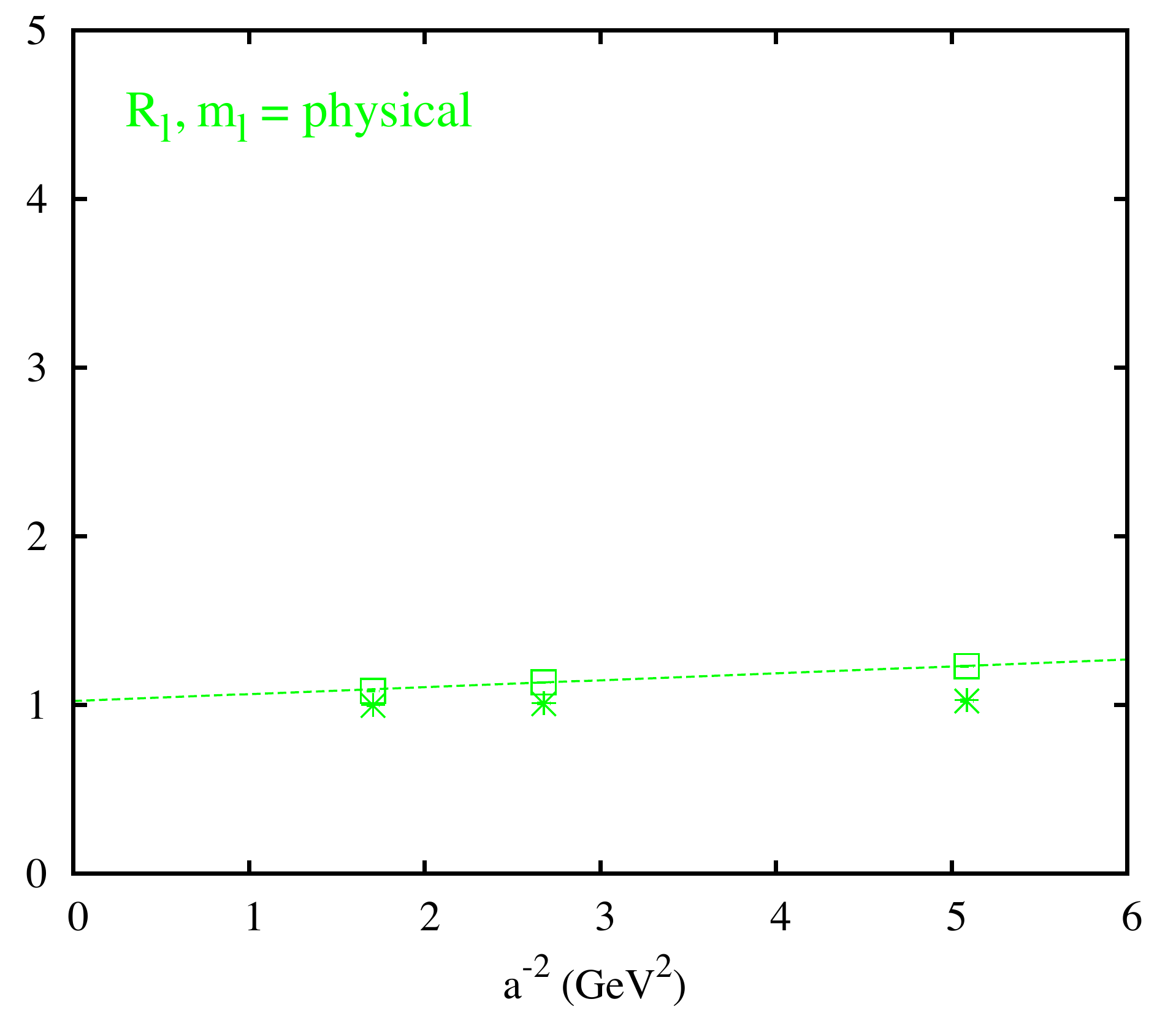}
\end{center}
\caption[kjhl]{ Results for $R_q$, defined in Eq.~\ref{eq:ratdef}, as a function of 
the square of the inverse lattice spacing. The plot on the left is for 
$s$ quarks and on the right for light quarks at the physical light quark 
mass ($m_l = (m_u+m_d)/2$). Open squares give raw results and pluses 
have tree-level perturbative corrections. Crosses have in addition one-loop 
perturbative corrections, using $\alpha_V^{(n_f=4)}(2/a)$. Dashed lines 
give a simple linear fit to the raw results.  
 }
\label{fig:raw}
\end{figure}

\section{Results}
We use gluon field configurations generated by the MILC collaboration 
including $u$, $d$ ($m_u=m_d$ here), $s$ and $c$ quarks in 
the sea at 3 different 
values of the lattice spacing and having $u/d$ quark masses down to 
the physical value~\cite{milc}. The quark formalism used is the Highly Improved 
Staggered Quark (HISQ) action~\cite{hisq} that has good chiral properties and small 
discretisation errors. 

Figure~\ref{fig:raw} shows our raw and corrected lattice results 
for $s$ quarks and for light quarks of physical mass. The quantity 
plotted is the ratio $R_q$ given by: 
\begin{equation}
R_q = - \frac{4m_q\langle \overline{\psi}\psi_q \rangle}{f^2_{\eta_q}M^2_{\eta_q}}.
\label{eq:ratdef}
\end{equation} 
Here $\eta_q$ is the pseudoscalar meson made from quarks of mass $m_q$. This 
is the $\pi$ for light quarks and a fictitious particle called the 
$\eta_s$ for $s$ quarks. We do not include disconnected diagrams 
in the $\eta_s$ correlator and so it is unable to decay to gluons 
or to mix with the physical 
$\eta$ and $\eta^{\prime}$ mesons. Because it has no valence light quarks, 
yet is not a heavy meson, 
it is a very useful particle in the determination 
of the lattice spacing~\cite{lat}. Its mass and decay constant in the 
continuum and chiral limits can be determined accurately in 
terms of those of the $K$ and the $\pi$~\cite{ups}. Interestingly, values very 
close to those expected in leading order chiral perturbation theory are 
found. 
$R_q$ is a good quantity to use in this analysis because it
has reduced finite volume and quark mass dependence over the 
condensate alone and is 
1 for quarks of zero mass, as given by the GMOR relation.

In Figure~\ref{fig:raw} the open squares give the raw results 
from the lattice, determining the condensate from the trace 
of the quark propagator. The $x$-axis is the inverse square of the 
lattice spacing and it is quite clear from these plots that there 
is a quark-mass dependent $1/a^2$ divergence. The pluses give 
the results after subtraction of the difference of the tree-level  
lattice QCD (for HISQ quarks) and $\overline{MS}$ perturbative expressions. 
The divergence is largely, but not entirely, removed. The crosses 
continue this process by removing the perturbative contribution 
through $\mathcal{O}(\alpha_s)$. This makes little difference because 
the coefficient of the $\mathcal{O}(\alpha_s)$ divergence is very small.  
It is then clear that the remaining divergence evident in the 
lattice calculations must come from higher orders in perturbation theory, 
and we take account of this in our fits. We also use the fact that the form 
of the divergence is strongly constrained and fit simultaneously results 
at multiple values of the quark mass and lattice spacing. 

\begin{figure}
\includegraphics[width=0.5\hsize]{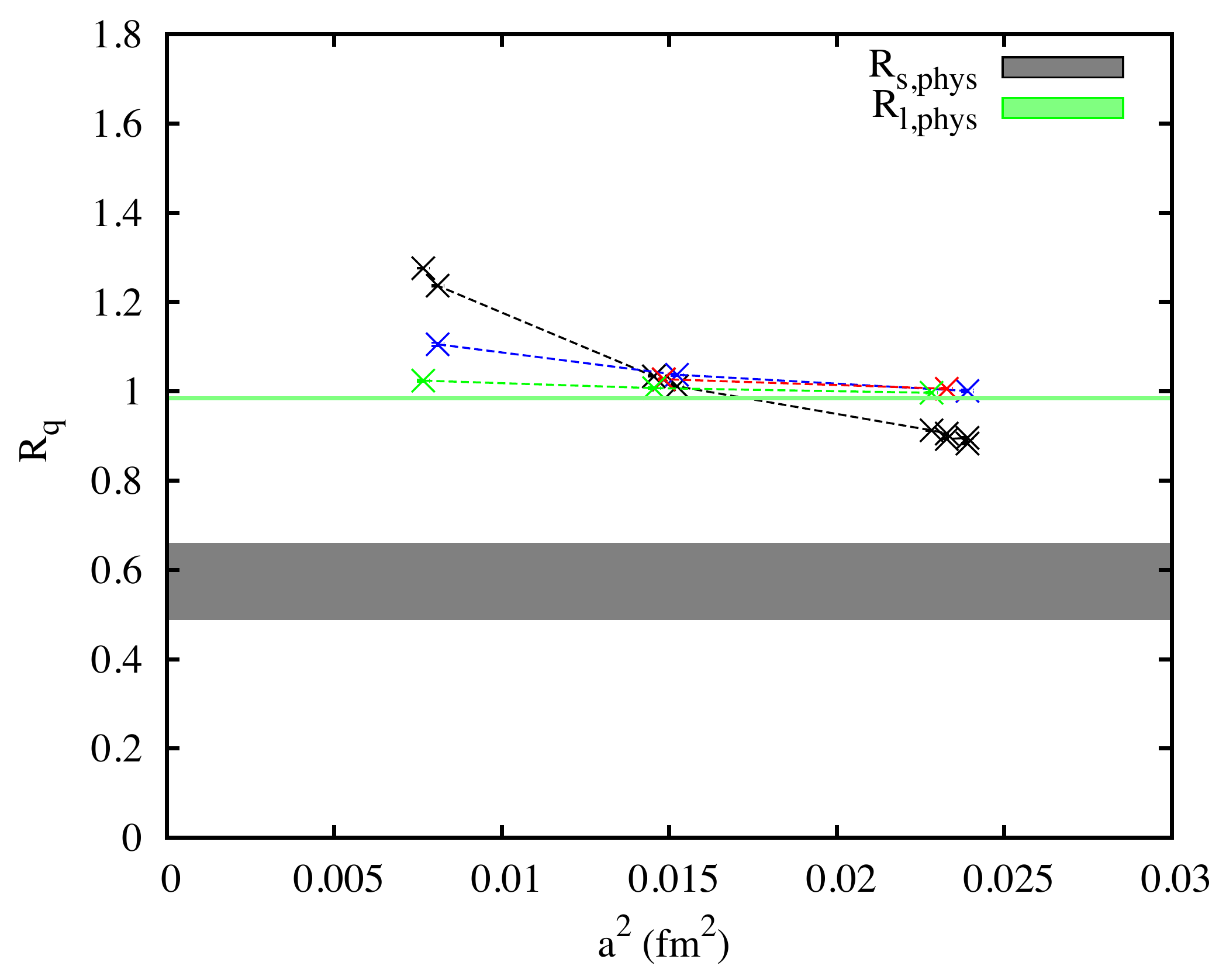}
\includegraphics[width=0.5\hsize]{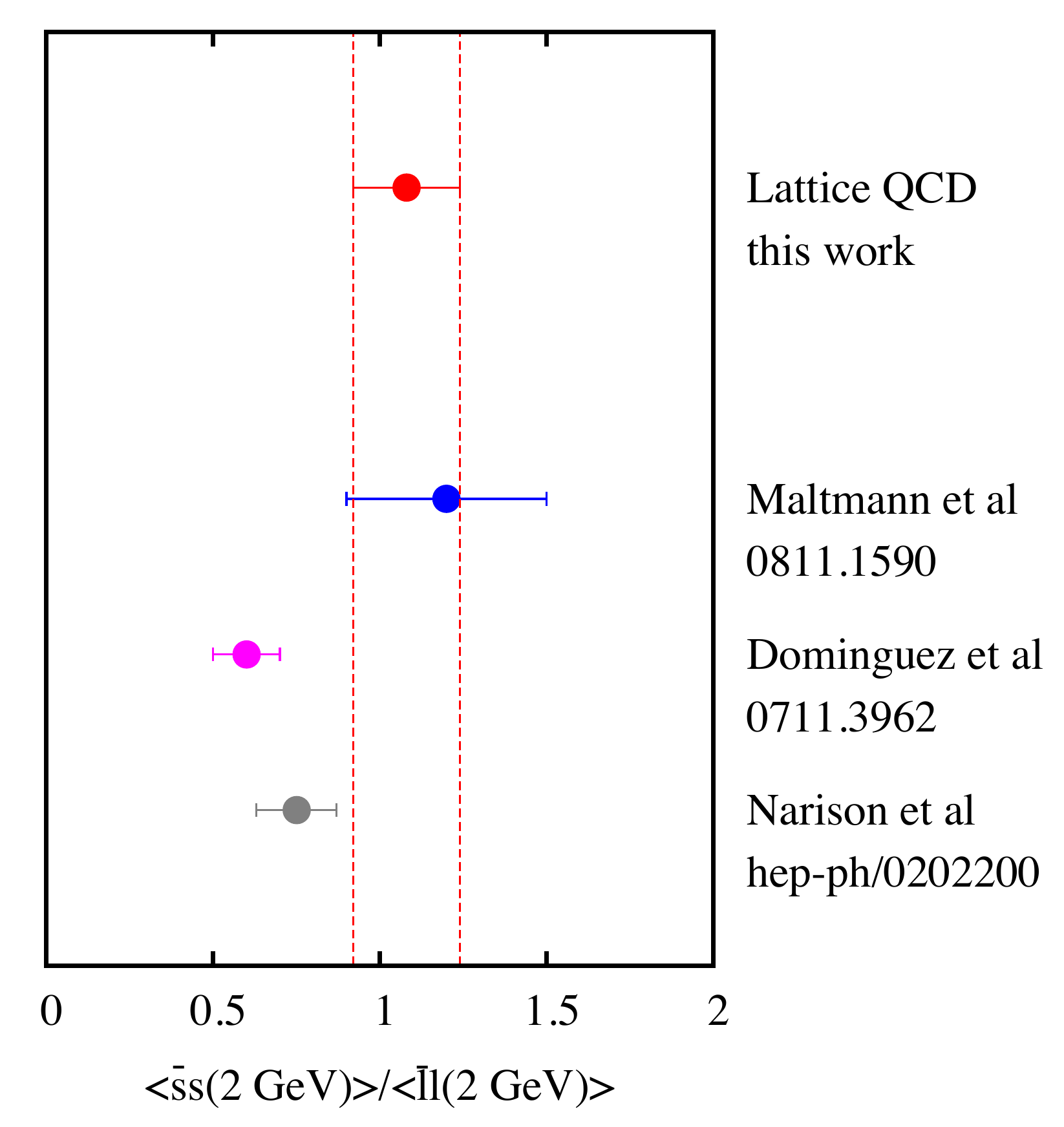}
\caption{ The figure on the left shows the results from fitting the ratio 
$R_q$ given by the crosses after subtracting the 
perturbative contribution through 
$\mathcal{O}(\alpha_s)$. 4 different quark masses are used: 
$m_s$ (black), $m_s/5$ (blue), 
$m_s/10$ (red) and the physical light 
quark mass (green). The green and black bands show the physical 
result from the fit with $\pm 1\sigma$ errors. Black is for the 
strange quark condensate and green for the light quark condensate. 
The figure on the right shows the comparison of our result for 
the ratio of strange to light quark condensate to those of previous 
sum rule analyses.
 }
\label{fig:fits}
\end{figure}

Figure~\ref{fig:fits} on the left shows 
the one-loop subtracted results from Figure~\ref{fig:raw} along with 
the final results from our fit.  
The $x$-axis is now the square of the lattice spacing and the grey 
and green bands show the continuum and chiral limit values of the 
strange and light quark condensates respectively, in the $\overline{MS}$ scheme 
at 2 GeV. Our fit form allows for discretisation errors and 
quark mass dependence of the physical condensate as well as the 
divergent pieces discussed above. Fitting the remaining divergence 
is the key issue in determining the final error, however, and the 
reason why the error on the strange quark condensate is 15\% and on 
the light quark condensate 0.5\%. Full details, along with an 
error budget, are given in~\cite{cond}. 

Our results are: 
\begin{eqnarray}
\langle \overline{s}s \rangle^{\overline{MS}}(2 \,\mathrm{GeV}) &=& -0.0245(37)(3) \,\mathrm{GeV}^3  
= -(290(15) \,\mathrm{MeV})^3 \nonumber \\
\langle \overline{l}l \rangle^{\overline{MS}}(2 \,\mathrm{GeV}) &=& -0.0227(1)(4) \,\mathrm{GeV}^3 
= -(283(2) \,\mathrm{MeV})^3 \nonumber \\
\frac{\langle \overline{s}{s} \rangle^{\overline{MS}}(2 \,\mathrm{GeV})}{\langle \overline{l}l \rangle^{\overline{MS}}(2 \,\mathrm{GeV})} &=& 1.08(16)(1) ,
\label{eq:ppval}
\end{eqnarray}
where the first error comes from our fitted result for $R_q$ 
and the second error in each case comes 
from the error in the quark masses, taken from lattice QCD results~\cite{mass}. 

Figure~\ref{fig:fits} shows on the right a comparison of our result 
for the ratio of the strange to light quark condensate to previous 
values from QCD sum rule calculations~\cite{maltman, domin, narison}. 
Our result has the advantage 
of being a direct determination with a full error budget. Our value 
for the strange quark condensate agrees well with independent results 
on gluon field configurations including $u$, $d$ and $s$ quarks in 
the sea and covering a wider range of values of $a$~\cite{cond}.

\end{document}